\documentclass[12pt]{article}
\usepackage{epsfig}
\usepackage{amsfonts}
\usepackage{latexsym}
\usepackage{amsmath}
\usepackage{mathrsfs}
\usepackage{hyperref}
\usepackage{wasysym}
\usepackage{tikz}
\usepackage{caption}
\usepackage{subcaption}
\usepackage{float}
\numberwithin{equation}{section}

\DeclareFontFamily{OT1}{rsfs}{}
\DeclareFontShape{OT1}{rsfs}{m}{n}{
<-7> rsfs5 <7-10> rsfs7 <10-> rsfs10}{}
\DeclareMathAlphabet{\mycal}{OT1}{rsfs}{m}{n}

\newcommand{\bea}{\begin{eqnarray}}
\newcommand{\eea}{\end{eqnarray}}
\def\non{\nonumber}
\def\pa{\partial}

\def\k{\kappa}

\def\e{\epsilon}

\def\p{\Phi}
\def\D{\Delta}
\def\non{\nonumber}

\def\b{\beta}

\def\tv{\tilde{v}}
\def\ads2{$\mathrm{AdS_2}$}
\def\nads2{$\mathrm{NAdS_2}$}
\def\sl2r{$\mathrm{SL(2,R)}$}
\def\tr{\tilde{R}}
\def\tt{\tilde{T}}
\def\tv{\tilde{V}}

\begin{document}

\begin{titlepage}
\unitlength = 1mm
\begin{center}

\setcounter{tocdepth}{2}


{ \LARGE{\textsc{Near-extremal black holes at late times, \\ \vspace{5mm} backreacted}}}

\vspace{0.8cm}
Shahar Hadar

\vspace{1cm}

{\it 
Center for the Fundamental Laws of Nature, Harvard University, \\ \vspace{0.1cm} Cambridge, MA 02138, USA \\ \vspace{0.3cm} and \\ \vspace{0.3cm} Max Planck Institute for Gravitational Physics (Albert Einstein Institute), \\\vspace{0.1cm}   Am M{\"u}hlenberg 1, 14476 Potsdam-Golm, Germany
\vspace{0.3cm}
}

\begin{abstract}
Black holes display universal behavior near extremality. One such feature is the late-time blowup of derivatives of linearized perturbations across the horizon. For generic initial data, this instability is regulated by backreaction, and the final state is a near-extremal black hole. The aim of this paper is to study the late time behavior of such black holes analytically using the weakly broken conformal symmetry of their near-horizon region. In particular, gravitational backreaction is accounted for within the Jackiw-Teitelboim model for near-horizon, near-extremal dynamics coupled to bulk matter.
\end{abstract}

\vspace{1.0cm}

\end{center}
\end{titlepage}

\pagestyle{plain}
\setcounter{page}{1}
\newcounter{bean}
\baselineskip18pt


\setcounter{tocdepth}{2}


\tableofcontents

\section{Introduction}\label{Introduction}

Black holes (BHs) in the limit of vanishing Hawking temperature exhibit behavior reminiscent of critical phenomena, and this has surfaced in various different contexts in BH physics, e.g. \cite{Hartnoll:2009sz},\cite{Gralla:2016jfc},\cite{Gralla:2018xzo}. In many-body systems near criticality, such behavior is explained by the emergence of a conformal symmetry. In the gravitational context, the BH's surface gravity vanishes in the extremal limit and the conformal symmetry is geometrically realized near the horizon \cite{Maldacena:1998uz},\cite{Guica:2008mu}. 
Such BHs are called near-extremal since they nearly saturate bounds on certain combinations of the mass, angular momentum and charges, and they bear particular significance, both theoretically and observationally. Astrophysically, neutral BHs with angular momentum close to the Kerr bound are believed to be common in the sky, and some possible observational signatures of these high-spin BHs, relevant for both gravitational-wave and optical observatories, have recently been proposed, for example \cite{Hadar:2014dpa},\cite{Gralla:2016qfw},\cite{Gralla:2017ufe},\cite{Gates:2018hub}. 

There has been much recent progress on understanding the late-time behavior of perturbed extremal BHs. In particular, it was discovered by Aretakis in a series of works \cite{Aretakis:2011ha},\cite{Aretakis:2011hc},\cite{Aretakis:2012ei},\cite{Aretakis:2012bm} that linearized perturbations of a massless scalar field on an extremal Reissner-Nordstr{\"o}m (RN) BH background display both stable and unstable properties: perturbations outside the horizon decay at late times, but on the horizon transverse derivatives generally do not decay and may even blow up. This behavior is closely related to Aretakis' discovery of an infinite number of conserved quantities on the horizon, one per mode $(\ell,m)$ on the transverse sphere, or conversely a conserved function on the sphere. The spherically symmetric mode $\phi_0$ of such perturbations, for example, behaves at late times ($v \to \infty$) on the horizon as
\bea
\partial_r^k \phi_0 \sim v^{k-1} \, ,
\label{aretakis behavior intro}
\eea
for smooth initial data supported on the future horizon, where $\{v,r\}$ are ingoing Eddington-Finkelstein coordinates. Several subsequent works on the subject, using different techniques, studied late-time behavior on the horizon for different types of extremal BHs and different types of perturbation fields \cite{Lucietti:2012xr},\cite{Casals:2016mel},\cite{Gralla:2016sxp},\cite{Zimmerman:2016qtn}, and in particular in \cite{Lucietti:2012sf} it was shown that a massless scalar displays Aretakis behavior on \emph{any} extreme BH background. The picture arising from these studies is that the Aretakis behavior is a general feature of extremal BHs, in the following sense: the $k$-th transverse-to-the-horizon derivative of the perturbation, at late times on the horizon, behaves (no worse than\footnote{\label{skip footnote}in fact, for generic initial data, (\ref{aretakis behavior intro-general}) is the precise decay/blowup rate modulo the so-called ``skip phenomenon'' \cite{Aretakis:2010gd},\cite{Angelopoulos:2018uwb}, which in some cases of integer $\D_-$, for some values of $k$, suppresses (\ref{aretakis behavior intro-general}) by an additional factor of $v^{-1}$---causing faster decay or slower blowup.}) 
\bea
\sim v^{k+\D_-} \, ,
\label{aretakis behavior intro-general}
\eea
where $\D_-$ is a negative $k$-independent, fixed non-universal constant.\footnote{which depends on: which mode is being considered (say in a spherical/spheroidal harmonic decomposition), the field's spin, whether or not the initial data has support on the horizon and the type of extremal BH under study.}

A convenient way to repackage (\ref{aretakis behavior intro-general}) (assuming an expansion in the radial coordinate) is to use the following ansatz for the field (assume a scalar field for simplicity):
\bea
\phi \simeq v^{\D_-} F\left[v \, (r-M)\right] \, ,
\label{aretakis behavior ansatz}
\eea
where $F(z)$ is an arbitrary function which is smooth at $z=0$. It is important to note that this type of behavior is consistent with the results of \cite{Angelopoulos:2018uwb}, which showed that for generic initial data in extremal RN the decay and blowup rates (\ref{aretakis behavior intro-general}) are saturated (again, modulo the skip phenomenon described in footnote \ref{skip footnote}).
Following \cite{Hadar:2017ven} (see also \cite{Gralla:2018xzo}), plugging (\ref{aretakis behavior ansatz}) into the 1+1 dimensional wave equation for each mode and considering the near-horizon, late-time limit reduces it to an ordinary differential equation for $F(z)$, which yields a \emph{universal} behavior for the field at late times near the horizon. For example, for the spherically symmetric mode $(\D_-=-1)$ of the above discussed neutral massless scalar around extreme RN:\footnote{Another universal piece $\sim 1/v$ arises in the case of non-vanishing $\ell=0$ Newman-Penrose constant; here we assume it does vanish.}
\bea
\phi_0 \simeq \frac{H_0}{v \left(2+v \, (r-M)/M^2\right)} \, ,
\label{universal late time solution}
\eea
at late times, where $H_0$ is interpreted as the $\ell=0$ Aretakis constant.

Since the Aretakis analysis concerns linearized perturbations and involves blowup, one must ask whether and how \emph{gravitational backreaction} becomes important. This was addressed in \cite{Murata:2013daa} numerically for spherically symmetric perturbations of a massless, neutral scalar field. The conclusions were that the instability persists under backreaction, and that for generic initial data, the BH approaches a new, \emph{non-extremal} solution at late times, and this allows perturbations to exponentially decay at times of the order of the inverse temperature. This means that even if the BH is perturbed with a small amplitude, certain quantities (with enough radial derivatives) can grow to parametrically larger values, and in particular not remain small, before eventually decaying.\footnote{Note also that, as shown in \cite{Murata:2013daa}, one may fine-tune the initial data such that the BH remains extremal, and the instability continues to evolve forever.}
This makes sense from an outside observer's point of view: energy is thrown down the hole, and this can raise the (effective) temperature, allowing dissipation; a picture which connects well also with \cite{Gralla:2016sxp} which studied the ``transient instability'' behavior of linearized perturbations to near-extremal Kerr BHs, finding Aretakis-like growth at intermediate times and eventually exponential decay at later times. 

The universal behavior of extreme BHs has a unifying explanation, and as stated above, in gravity as in many-body systems, it is the emergence of a conformal symmetry at criticality. From the gravitational point of view it is realized geometrically, and this can be seen by zooming in on the near-horizon region, where a large degree of symmetry is found: a global \sl2r isometry which is, at least in some cases, understood to be enhanced to a local infinite-dimensional asymptotic symmetry group \cite{Maldacena:1998uz},\cite{Guica:2008mu},\cite{Maldacena:2016upp}. There is a strong connection between the behavior discovered by Aretakis and this symmetry: the \sl2r isometry alone is enough to show that (\ref{aretakis behavior intro-general}) had to be the late-time behavior. This was argued in \cite{Lucietti:2012xr} for scalar perturbations in \ads2, then elaborated upon and extended to nonaxisymmetric modes with nonvanishing Aretakis constants in extremal Kerr in \cite{Hadar:2017ven}, and further generalized in \cite{Gralla:2018xzo}.

For near-extremal BHs, the conformal symmetry is broken, but only weakly. This was understood recently in a very different context, that of the Sachdev-Ye-Kitaev (SYK) model \cite{Kitaev talk},\cite{Sachdev:1992fk}, a certain 0+1 dimensional many-body fermionic quantum mechanics with all-to-all coupling, which is solvable in the limit of a large number of degrees of freedom. SYK at low energies has substantial overlap with a seemingly very different theory---the Jackiw-Teitelboim (JT) model \cite{Jackiw:1984je},\cite{Teitelboim:1983ux}, a 1+1 dimensional theory of dilaton gravity. As part of this line of work, in \cite{Maldacena:2016upp} the symmetry breaking pattern of JT and its corresponding near-\ads2 solutions were understood. In particular it was shown that the conformal symmetry, which is exact in the extreme limit, is weakly broken for near-extreme BHs; both explicitly by the small temperature and spontaneously by the choice of reparametrization relating the near-horizon time coordinate to the boundary (proper) time, and this structure practically determines the low-energy effective action (see also \cite{Jensen:2016pah},\cite{Engelsoy:2016xyb} for derivations of the low-energy `Schwarzian' action). Moreover, and importantly for this work, it was argued in \cite{Almheiri:2014cka}, \cite{Maldacena:2016upp} and shown in detail in some specific cases \cite{Nayak:2018qej},\cite{Moitra:2018jqs},\cite{Sarosi:2017ykf} that the JT model provides a universal description of the near-horizon gravitational dynamics of near-extremal BHs, via dimensional reduction.

In this paper, the late-time dynamics of near-extremal BHs is studied analytically, from the near-horizon perspective. The JT model allows for the description of backreaction effects while remaining in near-\ads2, as was pointed out in \cite{Almheiri:2014cka}. This enables one to both exploit the (weakly broken) symmetries, and to account for the out-of-equilibrium nature of the BH's evolution during the process; and this is an \emph{essential} effect to take into account if the perturbation's energy is comparable to the energy above extremality. From a technical point of view, a crucial point is that the backreaction can be monitored by promoting the boundary to a degree of freedom characterized by a worldline on \ads2; and the bulk fields, which propagate on the corresponding---locally \ads2---cutout geometry, couple to gravity only through a local interaction at the boundary. This simplified picture is used here to compute the late-time behavior of perturbations on the horizon and outside it. We study the simplest case of spherically symmetric scalar perturbations, which reduce to a massless scalar propagating on near-\ads2. For such perturbations with generic near-horizon initial data, we solve explicitly for the boundary degree of freedom describing the backreacted BH's evolution. Our analysis covers also the case of perturbations to the BH from afar. We then show how to use the above solution in order to infer the leading late-time behavior on and off the horizon. We show that perturbations decay exponentially at late times, with a decay constant determined by the final surface gravity, and this in turn may be inferred from the near-horizon picture. We further comment on how the above framework could be useful in analyzing more refined aspects of perturbations to such geometries.

We use a boundary-bulk or gravity-matter coupling at the level of the equations of motion which was derived in \cite{Maldacena:2016upp}. It could be interesting to carry out the computation with a coupling at the level of the action, which would presumably require to consider the two-sided BH (which should be equivalent to a Schwinger-Keldysh field doubling of the one-sided case). It could be interesting to compare the boundary behavior derived here to expectations from holography. Including backreaction means taking into account the variation in the BH's temperature during the evolution: a non-equilibrium process in a putative field theory dual, in which quenching pumps into the system an energy of the order of the initial thermal energy. Finally, it would be very interesting to work out in detail the analogous process in extreme Kerr and its corresponding near-NHEK geometry and spell out in detail the similarities and differences from the spherically symmetric charged case. 

In section \ref{Instability from symmetry}, we review the derivation of late time behavior at extremality, including the Aretakis instability and boundary decay from the near-horizon conformal symmetry. In section \ref{Transient behavior and quasinormal modes from symmetry} we derive in detail the near-extremal characrteristic behavior from the (weakly broken) symmetry, including the exponential decay at late time via quasinormal modes and transient behavior at intermediate times. Finally in section \ref{backreacted solution} we treat the backreacted late-time behavior from the near-horizon point of view using JT gravity.

\section{Extremal behavior from symmetry}\label{Instability from symmetry}

In this section we will review the argument that the decay and blow up rates of a massless scalar field and its derivatives on the horizon at late times follow from an analysis of the symmetries of the near-horizon region. We focus on this simple case for clarity, but generalizations will be commented upon. 

Begin with the Reissner-Nordstr{\"o}m solution\footnote{We use here gravitational units $G_N=c=1$. Note also that from here on, and differently from section \ref{Introduction}, hatted coordinates and quantities will be related to the full (extreme or near-extreme) RN geometry (\ref{RN solution}),(\ref{ERN solution}), while unhatted coordinates and quantities will be related to  near-horizon geometries (\ref{near horizon metric}),(\ref{nads2 solution}).}
\bea
ds^2 &=& -\left( 1-\frac{2M}{\hat{r}} + \frac{Q^2}{\hat{r}^2} \right) d\hat{t}^2 + \left( 1-\frac{2M}{\hat{r}} + \frac{Q^2}{\hat{r}^2} \right)^{-1} d\hat{r}^2 + \hat{r}^2 d\Omega^2  \, , \non \\
\hat{A}_{\hat{t}} &=& -\frac{Q}{\hat{r}} \, .
\label{RN solution}
\eea
Now define $\hat{\k} := \sqrt{1-Q^2/M^2}$, which describes the BH's deviation from extremality: near-extremal BHs are defined by $\hat{\k} \ll1$, and then $\hat{\k}$ is proportional to the BH's dimensionless surface gravity/temperature. Exactly extremal, zero temperature BHs satisfy $\hat{\k}=0$.

In this section we will consider the maximally electrically charged ERN solution, $\hat{\k}=0$ or $Q=M$,
\bea
ds^2 &=& -\left(1-\frac{M}{\hat{r}}\right)^2 d\hat{t}^2 + \left(1-\frac{M}{\hat{r}}\right)^{-2} d\hat{r}^2 + \hat{r}^2 d\Omega^2 \, , \non \\
A &=& -\frac{M}{\hat{r}} d\hat{t} \, ,
\label{ERN solution}
\eea
where $d\Omega^2$ is the line element on the 2-sphere.
A nondegenerate metric describing the near-horizon region can be obtained by introducing a 1-parameter family of coordinate and gauge transformations (see also recent discussion in \cite{Porfyriadis:2018yag},\cite{Porfyriadis:2018jlw})
\bea
R=\frac{\hat{r}-M}{\e  M} \, \, \, \, , \, \, \, \, T=\frac{\e \, \hat{t}}{M} \, \, \, \, , \, \, \, \, A=\hat{A}+d\hat{t}\, ,
\label{near horizon limit}
\eea
and taking the $\e \to 0$ limit. The resulting near-horizon solution is
\bea
\frac{ds^2}{M^2} &=& -R^2 dT^2 + R^{-2} dR^2 + d\Omega^2 \, , \label{near horizon metric}  \\ \non \\
\frac{A}{M} &=& R dT \, . \label{near horizongauge field}
\eea
The geometry (\ref{near horizon metric}) is referred to as the Robinson-Bertotti universe or $\mathrm{AdS_2\times S^2}$. In what follows, we will sometimes find it useful to work with different coordinate systems on its \ads2 factor. Introduce ingoing Eddington-Finkelstein coordinates $\{V,R\}$ by
\bea
R_*=\int\frac{dR}{R^2}  \, \, \, \, \, ; \, \, \, \, \, V:=T+R_* = T-1/R \, ,
\label{tortoise coordinate def}
\eea
where on the right hand side of the second equation above we made a specific choice for the integration constant. The metric then becomes
\bea
\frac{ds^2}{M^2} &=& -R^2 dV^2 + 2 dV dR + d\Omega^2 \, . \label{near-horizon metric in EF coordinates}  
\eea
Yet another useful coordinate system is the so-called double-null or lightcone coordinates $\{V,U\}$, where we take
\bea
 U:=T-R_* = T+1/R \, ,
\label{U coordinate def}
\eea
yielding the metric 
\bea
\frac{ds^2}{M^2} &=& - 4 \, \frac{ dV dU}{(V-U)^2} + d\Omega^2 \, . \label{near-horizon metric in double-null coordinates}  
\eea

The solution (\ref{near horizon metric}),(\ref{near-horizon metric in EF coordinates}) is invariant under the 1-parameter family of coordinate transformations (isometries)
\bea
T-T_\infty &=& -\frac{\tr^2 \tt}{\tr^2 \tt^2 -1} = -\frac{1+\tv\tr}{\tv(2+\tv\tr)} \, , \non \\
R &=& \frac{\tr^2 \tt^2 -1}{\tr} = \tv(2+\tv\tr)\, ,
\label{coordinate trans ads2->ads2}
\eea
where $\{\tt,\tr\}$,$\{\tv,\tr\}$ are the new coordinates with metrics (\ref{near horizon metric}), (\ref{near-horizon metric in EF coordinates}) respectively and $T_\infty$ is an arbitrary parameter.\footnote{Supplemented by the gauge transformation $A=\tilde{A}+d(-2\tanh^{-1}(\tr \,\tt))$, the electric field (\ref{near horizongauge field}) is preserved as well.} 
The transformation (\ref{coordinate trans ads2->ads2}) may be thought of as a composition of an (exponentiated/large) \ads2 global time translation and Poincar{\'e} time translation.

\begin{figure}[h!]
	\centering
	
	\resizebox{0.4\textwidth}{!}
	{
		
\begin{tikzpicture}
\draw[thick] (0,-2.5) -- (0, 8.5);
\draw[thick] (-4,-2.5) -- (-4, 8.5);

\draw[thick] (0,-2) to (-4, 2) to (0,6);

\draw[thick,dashed] (0,1.7) to (-3.15, 4.85);

\draw[thick] (0,0) to (-4, 4) to (0,8);

\draw (1,6.3) node[rotate=0]{$T=T_\infty$};
\draw (1,5.7) node[rotate=0]{$\tv\to\infty$};

\draw [thick,dotted] (0,6) circle (7pt);

\draw (-1.5, 5) node[rotate=45]{$\tr=0$};
\draw (+0.35, 1.75) node[rotate=90]{$\tr=\infty$};

\draw (-2.5, 6) node[rotate=45]{$R=0$};
\draw (+0.35, 3.75) node[rotate=90]{$R=\infty$};

\end{tikzpicture}

	}
	
	\caption{Penrose diagram displaying the isometry (\ref{coordinate trans ads2->ads2}). The coordinates $\{T,R\}$ cover the upper Poincar{\'e} patch with metric (\ref{near horizon metric}), while the coordinates $\{\tt,\tr\}$ cover the lower Poincar{\'e} patch. The dashed null line represents a possible initial data surface, and the dotted circle highlights the neighborhood of $\tv\to\infty$ on which we are focusing. This point is the future endpoint of the future horizon and of the boundary of the patch $\{\tt,\tr\}$.}
	\label{figure1}
\end{figure}
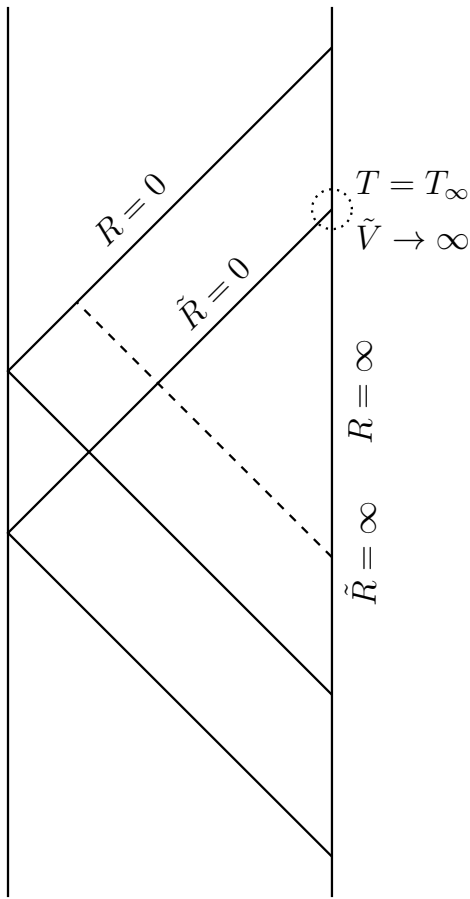

Importantly, even though (\ref{coordinate trans ads2->ads2}) is an isometry, and therefore preserves the metric locally, it may modify the physical situation under consideration since the two coordinate systems in (\ref{coordinate trans ads2->ads2}) cover different patches of the maximal analytic extension of this geometry (see figure \ref{figure1}), and in particular different portions of the boundary. Notably, this transformation will modify the physical situation whenever the $AdS_2$ spacetime is thought of as a near-horizon limit of a higher dimensional extremal BH, in which case the boundary is thought of as a surface upon which matching to the far-region of the extreme RN geometry is performed, and this defines a preferred choice of boundary.

Now consider a massless scalar $\phi$ propagating on this geometry. Its equation of motion is $\square \phi = 0$. Decomposing into spherical harmonics
\bea
\phi=\sum_{\ell m} \phi_{\ell m}(T,R) Y_{\ell m}(\Omega),
\label{spherical harmonic decomposition}
\eea
it can be written as
\bea
\left( R^{-2} \pa^2_T +\pa_R R^2 \pa_R -\ell(\ell+1)\right)\phi_{\ell m}(T,R)=0 \, .
\label{wave eqn modes}
\eea
Assuming separation of variables, or time dependence of the form $\phi \sim e^{-i \omega T}$, shows that for $R \to \infty$ the modes behave as $\phi_{\ell m \omega} \sim R^{\Delta_{\pm}}$, where
\bea
\Delta_{\pm} = -1/2 \pm (\ell+1/2) \,.
\label{deltas}
\eea

Let us first analyze the $\ell=0$ case. The d'Alambertian in \ads2 (\ref{wave eqn modes}) is proportional to the $1+1$ dimensional Minkowski space wave equation, written in terms of the spatial coordinate $1/R$,
\bea
\square  = (1/R)^{2} \left( -\partial^2_T +\partial^2_{(1/R)} \right)  \, ,
\label{box in ads2}
\eea
and therefore the general solution can be written as
\bea
\phi = f(U)+g(V) \, , 
\label{l=0 general solution}
\eea
a superposition of arbitrary rightmoving and leftmoving waves determined by the functions $f$, $g$ which we will assume are smooth. Using (\ref{coordinate trans ads2->ads2}) gives
\bea
\phi = f\left(T_\infty -\frac{\tr}{\tv \tr + 2}\right)+g\left(T_\infty - 1/\tv\right) \, .
\label{l=0 solution}
\eea
Imposing the physical requirement that the field decays at late times outside the horizon, $\phi \to 0$ for $\tt \to \infty$, we get $f(T_\infty) = - g(T_\infty)$. This means that, on the horizon as well, $\phi \to 0$ as $\tv \to \infty$. The behavior of radial derivatives on the horizon can be inferred by directly operating on (\ref{l=0 solution}) with $\pa^k_{\tr}$ for $k\geq1$, evaluating at $\tr=0$ and taking the $\tv\to\infty$ limit:
\bea
\left.\pa^k_{\tr} \phi \right|_{\mathrm{horizon}} \approx f^{(k)}(T_\infty) \, \frac{ (-1)^k \, k!} {2^k} \, \tv^{k-1} \, ,
\label{l=0 aretakis behavior}
\eea
where $f^{(k)}$ stands for the $k$-th derivative of $f$. Hence the Aretakis behavior is recovered from near-horizon conformal symmetry.


Now assume $\ell>0$. A general solution can be written for large $R$ as
\bea
\phi = f_+(T) \, R^{\Delta_+} \left(1+\mathcal{O}(R^{-1})\right)
 + f_-(T) \, R^{\Delta_-}\left(1+\mathcal{O}(R^{-1})\right)
 \, ,
\label{general superposition of modes}
\eea
for some smooth $f_{\pm}$.
Imagine we prescribe some initial data on, say, an ingoing null surface as depicted by the dashed line in figure \ref{figure1}. The behavior near the boundary at some future time $T=T_\infty$ will be given by the leading order piece (in $1/R$) of equation (\ref{general superposition of modes}). Now, as in the $\ell=0$ case, we can use the diffeomorphism (\ref{coordinate trans ads2->ads2}) to place the point $(T=T_\infty,R=\infty)$ (encircled in figure \ref{figure1} by the dotted line) at future timelike infinity of the patch described by the coordinates $\{\tv,\tr\}$. We will think of this patch as the physical one, i.e. that which describes the near-horizon region of a higher-dimensional extreme BH geometry. 

The late-time behavior is only affected by the boundary conditions close to $T_\infty$. In general they may be of mixed, Robin-type nature, and time-dependent, but we will \emph{assume} that they are close enough to Dirichlet conditions of $\phi|_{\mathrm{bdy}}=0$ at late times---in particular, that $\partial_T^k f_+(T_\infty)=0$ for $k<\Re{(\D_+)}-\Re{(\D_-)}=2\ell+1$. It makes sense to pick boundary conditions which satisfy this condition (a much weaker condition than, for example, simply putting Dirichlet $\phi|_\mathrm{bdy}=0$ boundary conditions for any time) since this assumption yields decay rates for the field itself (before discussing the behavior of derivatives) which are consistent with known results on full geometries, both on and off the horizon, as will be seen below in the boundary analysis. The behavior of derivatives is inferred by the symmetry argument presented here.\footnote{Note that for cases when there are modes with complex scaling dimensions $\D_\pm=1/2\pm i \xi$ with real $\xi$, as is sometimes the case for extreme Kerr or charged perturbations on extreme RN, the only required assumptions are that the field outside the horizon at late times decays and that the boundary conditions are given by some linear combination of the $\pm$ modes at late times.}

Assuming these boundary conditions and plugging (\ref{coordinate trans ads2->ads2}) into (\ref{general superposition of modes}), we find that for $\tv\to\infty$, $\tr \tv$ fixed, to leading order,
\bea
\phi =  \tv^{-\ell-1} \left[ - f_+^{(2\ell+1)}(T_\infty) \left(1+ \tv \tr \right)^{2\ell+1}  + f_-(T_\infty) \right] \left(  2+ \tv  \tr \right)^{-\ell-1}  \, , \non \\
\label{behavior from symmetry ads2}
\eea
which is of the form 
\bea
\sim \tv^{\Delta_-} g(\tr \tv) \, ,
\label{ansatz}
\eea
for a function $g(z)$ which is regular at $z=0$, just as in (\ref{aretakis behavior ansatz}).
Taking $k$ derivatives and evaluating at the horizon, therefore, one finds
\bea
\left. \pa_{\tr}^k  \phi \right|_{\mathrm{horizon}} \to {\tv}^{\Delta_-+k} \, \, \, \, \, as  \, \, \, \, \, \tv \to \infty \, ,
\label{Aretakis behavior}
\eea
the Aretakis decay/blow up rates for nonzero Aretakis constants. A similar argument also works more generally for perturbations around various extremal BHs, and diferent types of perturbations. Note that in the general case where $\Re{(\D)}_+-\Re{(\D_-)}$ is not an integer, the first term on the right hand side of equation (\ref{behavior from symmetry ads2}) will be absent.  

The significance of the above argument is that it gives insight into why the Aretakis behavior is a universal feature of extremal BHs. Moreover, under the assumption of the above boundary conditions, it gives a remarkably simple way to predict the decay/blow-up rates for any such BH: one only needs to determine the exponent $\Delta_-$ which is fixed by the large-$R$ limit of the near-horizon wave equation. The argument has been used in \cite{Hadar:2017ven} to derive for the first time the decay/blowup rates for extreme Kerr with nonzero Aretakis constants, yielding the a priori surprising result that for the ``worst-behaving'' non-axisymmetric modes (those with complex $\D_\pm$), the rates are the same as in the case with vanishing Aretakis constants. For fields of nonzero spin, at least when perturbations separate and decouple, as in e.g. the case of the Teukolsky equations, a similar argument should go through. In this case, the diffeomorphism we use will need to be supplemented by a tetrad rotation (c.f. \cite{Hadar:2014dpa}).

The boundary late-time behavior may also be addressed using a similar symmetry argument, under a similar (one $T$-derivative stronger, as seen below) assumption on the boundary conditions. It allows to recover in detail the known late-time behavior (``tail'') \cite{Ori:2013iua},\cite{Bhattacharjee:2018pqb} of fields outside the horizon in the full extreme RN geometry. Again, it is natural to expect that an analogous assumption on the boundary conditions in similar near-horizon geometries such as NHEK \cite{Hadar:2017ven}, will recover the ``tail'' behavior from the symmetry argument presented here, for more general extreme BHs. 

Using equation (\ref{coordinate trans ads2->ads2}) with $\tv \to \infty$, $\tv \tr \gg 1$, places us on the boundary at late times. Imposing the condition that on the boundary $\phi(T_\infty,R=\infty)=\pa_T \phi(T_\infty,R=\infty)=0$ (one $T$-derivative stronger than in the above analysis of horizon behavior, see also \cite{Lucietti:2012xr}), we can express the $\ell=0$ solution for $T \to T_\infty$ as
\bea
\phi \simeq \frac{2 f'(T_\infty)}{\tt^2 \tr} + \left(f^{(2)}(T_\infty)+g^{(2)}(T_\infty)\right) \, 1/\tt^2 \, .
\label{boundary late time behavior l=0 ads2 1}
\eea
The condition $\phi(T_\infty,R=\infty)=\pa_T \phi(T_\infty,R=\infty)=0$ enforces the decay rate (\ref{boundary late time behavior l=0 ads2 1}) and is what we regard in this case to be boundary conditions that approach Dirichlet, $\phi|_\mathrm{bdy}=0$, ``rapidly enough'' at future timelike infinity to yield the known result in extreme RN (\cite{Ori:2013iua}). For example, requiring $\phi|_\mathrm{bdy}=0$ precisely from some $T<T_\infty$ onwards, implies the above assumption. In such a case, or in fact whenever $\pa^2_T \phi(T_\infty,R=\infty)=0$ is also imposed, the second term on the RHS of (\ref{boundary late time behavior l=0 ads2 1}) will drop out and only the first one will survive.

For $\ell \geq 1$, we can use the form (\ref{general superposition of modes}) and plug (\ref{coordinate trans ads2->ads2}) into it. Taking the $\tr\to\infty$ limit yields, for large $\tt$,
\bea
\phi \sim \tr^{\Delta_-} \tt^{2 \Delta_-} \, ,
\label{boundary late time behavior l=0 ads2 2}
\eea
for boundary conditions which are close enough to Dirichlet $\phi|_{\mathrm{bdy}}=0$ at late times. In the extremal RN case considered here, (\ref{boundary late time behavior l=0 ads2 2}) is compatible with the result of \cite{Ori:2013iua}---namely that the late-time tails outside extreme RN behave as $\phi \sim \tt^{- (2 \ell+2)}$.

\section{Near-extremal behavior from symmetry} \label{Transient behavior and quasinormal modes from symmetry}

After reviewing the derivation of late-time behavior of linearized perturbations from a symmetry argument, in this section we will go further and show that the near-extremal behavior can also be deduced directly from the enhanced symmetry of the near-horizon geometry, which is only weakly broken for such BHs. We will recover the so called ``transient instability'' behavior \cite{Gralla:2016sxp} at intermediate times and the exponential decay at later times. Moreover, the decay outside the horizon at late times as a superposition of resonances with particular frequencies, or quasinormal modes (QNMs), will be recovered from the symmetry argument.

\begin{figure}[h!]
	\centering
	
	\resizebox{0.4\textwidth}{!}
	{
		
\begin{tikzpicture}
\draw[thick] (0,-1) -- (0, 9);
\draw[thick] (-4,-1) -- (-4, 9);

\draw[thick] (-3.5, 3.5) to (0,7);

\draw[thick,dashed] (0,1.7) to (-3.15, 4.85);

\draw[thick] (0,0) to (-4, 4) to (0,8);

\draw (1,7.3) node[rotate=0]{$T=T_\infty$};
\draw (1,6.7) node[rotate=0]{$v\to\infty$};

\draw [thick,dotted] (0,7) circle (7pt);

\draw (-1.3, 5.2) node[rotate=45]{$r=0$};
\draw (+0.35, 2.75) node[rotate=90]{$r=\infty$};

\draw (-2, 6.5) node[rotate=45]{$R=0$};
\draw (+0.35, 4.75) node[rotate=90]{$R=\infty$};

\end{tikzpicture}

	}
	
	\caption{Penrose diagram displaying the mapping (\ref{coordinate trans ads2->Nads2}). The coordinates $\{T,R\}$ cover the Poincar{\'e} patch $0<R<\infty$ with metric (\ref{near horizon metric}), while the coordinates $\{t,r\}$ cover the Rindler patch $0<r<\infty$ with metric (\ref{nads2 solution}). The dashed null line represents a possible initial data surface.}
	\label{figure2}
\end{figure}
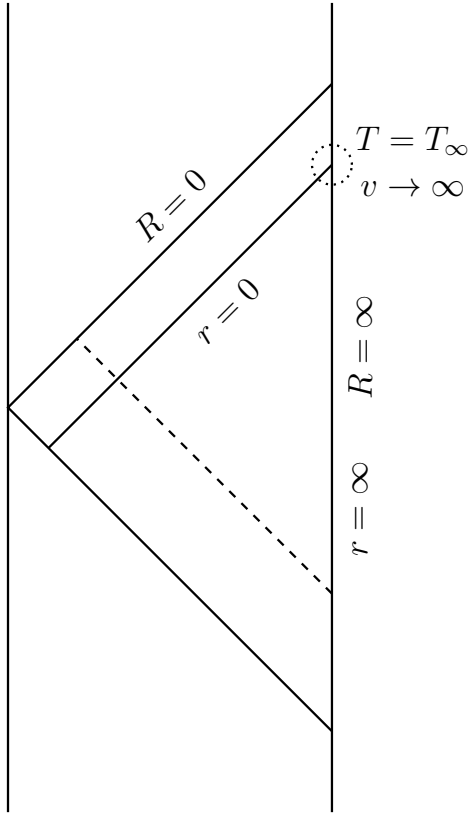

Consider a near-horizon limit of (\ref{RN solution}) as follows. Define a 1-parameter family of solutions, coordinate systems (and gauges) by taking (\ref{RN solution}) with $\hat{\k}=\e \kappa$, and
\bea
r=\frac{\hat{r}-r_+}{\e r_+} \, \, \, \, , \, \, \, \, t = \frac{\e \hat{t}}{M} \,  \, \, \, , \, \, \, \, a = \hat{A}+d\hat{t}  \, ,
\label{coordinate nads2}
\eea
where $r_+ = M (1+\hat{\kappa})$ is the location of the outer horizon, and take the $\e\to0$ limit. The resulting solution is
\bea
\frac{ds^2}{M^2} = -r(r+2\kappa) dt^2 + \frac{dr^2}{r(r+2\kappa)} + d\Omega^2   \, \, \, \, , \, \, \, \, a = M(r+\kappa)dt \, .
\label{nads2 solution}
\eea
This metric is sometimes referred to as the near-\ads2 metric. A fact we will use below is that even though the metrics (\ref{nads2 solution}) and (\ref{near horizon metric}) describe the same geometry locally, they cover different patches of the maximally extended geometry (which would be conveniently covered by global \ads2 coordinates). An explicit example for a mapping between a patch described by (\ref{nads2 solution}) into a part of a patch described by (\ref{near horizon metric}) is the family of large diffeomorphisms (see figure \ref{figure2} for a corresponding Penrose diagram)\footnote{One can supplement this coordinate transformation with a gauge transformation $a=A+d\left( \,\ln\sqrt{1+2\kappa/r} \, \,\right)$ in order that the electric field in (\ref{near horizongauge field}) transforms into the form in (\ref{nads2 solution}).}
\bea
T-T_\infty &=& - e^{-\k t}\frac{r+\k}{\sqrt{r(r+2\k)}} \, , \non \\
R &=& \frac{1}{\k} e^{\k t} \sqrt{r(r+2\k)} \, ,
\label{coordinate trans ads2->Nads2}
\eea


We will now use this mapping in order to derive the late-time behavior of scalar perturbations on the future horizon of near-extremal BHs. 
The transformation (\ref{coordinate trans ads2->Nads2}) places the point $\{T=T_\infty,R=\infty\}$ at the future boundary of both the future horizon and the boundary $r\to\infty$. Defining a tortoise coordinate
\bea
r_* = \frac{1}{2\kappa} \ln \frac{r}{r+2\kappa} \, ,
\label{tortoise coord nads2}
\eea
and an outgoing Eddington-Finkelstein coordinate $v:=t+r_*$, we can write
\bea
V &=& -e^{-\kappa v}+T_\infty \, , \non \\
U &=& -e^{-\kappa v} \frac{r}{r+2\kappa}+T_\infty \, .
\label{approx near future point}
\eea

Consider first the $\ell=0$ case and a solution of the form (\ref{l=0 general solution}). Similarly to (\ref{l=0 solution}), the solution here can be written as
\bea
\phi = f\left(T_\infty -e^{-\kappa v} \frac{r}{r + 2\kappa}\right)+g\left(T_\infty -e^{-\kappa v} \right) \, ,
\label{l=0 solution nads2}
\eea
where $f$, $g$ are arbitrary smooth functions. Expanding around $T_\infty$ gives
\bea
\phi = \sum_{n=1}^{\infty} (-)^{n} e^{- n \kappa v} \left[ f^{(n)}(T_\infty) \left(\frac{r}{r+2\kappa}\right)^n +g^{(n)} (T_\infty) \right] \, .
\label{nads2 qnms l=0}
\eea
The relation between $f^{(n)}$ and $g^{(n)}$ is determined by the boundary conditions. Specifically, since we would like the field to decay on the boundary at late times we set $f^{(0)}(T_\infty)+g^{(0)}(T_\infty)=0$ and therefore have omitted the $n=0$ element of the sum.

Expression (\ref{nads2 qnms l=0}) describes a familiar behavior. For any value of the radial coordinate, including the horizon $r=0$ and the boundary $r=\infty$, it is a sum over decaying exponentials. These describe nothing but a superposition of \emph{quasinormal modes (QNMs)}. We identify the QNM frequencies to be
\bea
\omega_n = -i \kappa n \, ,
\label{near horizon QNM frequencies}
\eea
or in terms of far-region frequency, i.e. that which is conjugate to the far-region time coordinate $\hat{t}$,
\bea
\hat{\omega}_n = -i \hat{\k} n \, .
\label{far region QNM frequencies}
\eea
As mentioned above, the $n=0$ term is absent, by boundary conditions, which means that $n=1,2,\ldots$.\footnote{The $n=0$ mode is redundant in this context. It is a zero mode with no spatial or time dependence---and simply corresponds to the shift symmetry $\phi \to \phi+\mathrm{const.}$ of the spherically symmetric sector of the field we are considering.} The QNM's amplitudes are the coefficients in front of the decaying exponentials in (\ref{nads2 qnms l=0}). This constitutes a derivation of $\ell=0$ near-extremal QNMs directly from the near-horizon conformal symmetry. 

For $\ell>0$, we will use
\bea
T-T_\infty &=& - e^{-\k v} \, \, \frac{r+\k}{r+2\k} \, , \non \\ \non \\
R &=& \, e^{\k v} \, \, \frac{r+2\k}{\k} \, ,  \\ \non
\label{bdy limit nads2}
\eea
and the expansion near the boundary point $\{T=T_\infty,R=\infty\}$,
\bea
\phi = f_+(T) \, R^{\Delta_+} \left(1+\sum_{k=1}^{\infty}a_k R^{-k}\right)
 + f_-(T) \, R^{\Delta_-} \left(1+\sum_{k=1}^{\infty}b_k R^{-k}\right)
 \, ,
\label{l>0 solution nads2}
\eea
where we assume that $f_\pm$ are smooth at $T=T_\infty$ and may be expanded near $T=T_\infty$ as
\bea
f_{\pm}(T) = \sum_{n=0}^{\infty} f_{\pm}^{(n)}(T_\infty) (T-T_\infty)^n \, .
\label{taylor expansion fpm}
\eea
Note that the sums in the brackets in (\ref{l>0 solution nads2}) do not contain logarithmic terms \cite{Skenderis:2002wp}.

Now, let us for simplicity assume that 
\bea
f^{(n)}_+(T_\infty) = 0  \, \, \, \,  ,  \, \, \, \, \forall \, n \, ,
\label{bdy bdy condition nads2}
\eea
enforcing boundary conditions which are ``close enough'' to Dirichlet, $\phi|_\mathrm{bdy}=0$, at late times (this condition can be relaxed to an assumption on only a finite number of derivatives, as in section \ref{Instability from symmetry}).
We can now rewrite (\ref{l>0 solution nads2}) as
\bea
\phi=\sum_{n=1}^{\infty} \, A_{n \ell} \, \,  e^{-\kappa (-\Delta_-+n-1) v} \, ,
\label{nads2 qnms l>0}
\eea
with
\bea
A_{n \ell}:=\sum_{j=0}^{n-1} f_-^{(j)}(T_\infty) \,\, b_{n-1-j} \,\, (-1)^{j} \,\,  \frac{\kappa^{n-1-j-\Delta_-} \,\, (r+\kappa)^{j}}{(r+2\kappa)^{n-1-\Delta_-}} \, .
\label{nads2 qnm amps l>0}
\eea
Similarly to the $\ell=0$ case, we interpret this expression as a sum over QNMs with frequencies
\bea
\omega_{n \ell} = -i \kappa (-1-\Delta_-+n) = -i \kappa (\ell+n)\, ,
\label{near horizon QNM frequencies l>0}
\eea
or far region frequency
\bea
\hat{\omega}_{n \ell} = -i \hat{\k} (-1-\Delta_-+n) = -i \hat{\k} (\ell+n) \, .
\label{far region QNM frequencies l>0}
\eea
for $n=1,2,\ldots$.
The boundary behavior can be extracted by taking the $r \to \infty$ limit in (\ref{nads2 qnms l>0}).

Note that the above family of QNMs (\ref{far region QNM frequencies l>0}) are usually referred to as ``weakley damped QNMs''. These and related modes have been extensively discussed recently (for example, in \cite{Dias:2018ufh}) following the suggestion of \cite{Cardoso:2017soq} that their existence in RN-de Sitter spacetimes implies a violation of the strong cosmic censorship conjecture. Note also that matching to the far region will determine the form of the QNM wavefunctions (\ref{nads2 qnm amps l>0}) but will not modify (to leading order in $\hat{\k}$) the frequencies (\ref{far region QNM frequencies l>0}).

It is natural to ask whether the Aretakis instability occurs for precisely extremal BHs only, or if it has some trace in the near-extremal limit. This was considered in \cite{Murata:2013daa} within an analytical ``toy model'' attempting to explain the behavior seen in the full nonlinear numerical evolution, and in \cite{Gralla:2016sxp}, for extremal Kerr with vanishing Aretakis constants, where a frequency-domain analysis was used. These analyses find qualitatively similar behavior---a ``transient instability'': for
moderately late times $1\ll \hat{v} \ll M \hat{\k}^{-1}$ an Aretakis-like behavior with power-law blow-ups, while for later times $\hat{v} > M \hat{\k}^{-1}$ the exponential decay (\ref{nads2 qnms l=0}), (\ref{nads2 qnms l>0}) is restored. Here we will show, again using a large diffeomorphism, that this behavior too can be recovered using the symmetry argument.

Consider the boundary to bulk retarded propagator $G(V,R;V')$ in \ads2 with Dirichlet $\phi|_\mathrm{bdy}=0$ boundary conditions. It constitutes a solution to the wave equation $\square \phi = 0$ that arises from turning on an instantaneous source term on the boundary, at some time that we will arbitrarily take to be $V'=0$. This propagator was recently computed in \cite{Gralla:2018xzo}. It is given, in ingoing Eddingtton-Finkelstein coordinates, by
\bea
G(V,R;0) = a_0 \, \Theta(V) \, V^{\D_-} \, \left( 2+V R \right)^{\D_-} \, ,
\label{boundary to bulk propagator}
\eea
where $a_0$ is a constant amplitude and $\Theta(V)$ is the Heaviside step function which enforces causality, or equivalently the condition of no incoming radiation from the past horizon. Note that (\ref{boundary to bulk propagator}) is of the universal form (\ref{ansatz}).

Using (\ref{coordinate trans ads2->Nads2}), (\ref{approx near future point}), we write
\bea
V &=& T_\infty \, (1-e^{-\k(v-v_0)}) \, , \non \\
R &=& T^{-1}_\infty \, \, \frac{r+2\k}{\k} \, e^{\k(v-v_0)} \, ,
\label{V and R}
\eea
where $T_\infty \equiv e^{-\k v_0}$. Using the mapping (\ref{V and R}) gives 
\bea
G(v,r;v_0) = a_0 \, T_\infty^{\D_-} \, \Theta(v-v_0) \, ( 1-e^{-\k(v-v_0)})^{\D_-} \left( 2+(e^{\k(v-v_0)}-1) \, \frac{r+2\k}{\k}  \right)^{\D_-} \, .
\label{boundary to bulk propagator near ads2}
\eea
This is the boundary to bulk propagator in near-\ads2, obtained from the \ads2 propagator by symmetry. Note that the no-incoming wave and boundary conditions are preserved by the transformation, and that here in the new near-\ads2 coordinates the source is turned on at $v=v_0=(-1/\kappa) \ln T_\infty $.
For $0<\k (v-v_0) \ll 1$, we can expand (\ref{boundary to bulk propagator near ads2}) to leading order in $\k (v-v_0)$ and obtain 
\bea
G(v,r;v_0) \approx a_0 \, (\k T_\infty)^{\D_-} \, (v-v_0)^{\D_-} \left( 2+ (v-v_0) \, (r+2\k)  \right)^{\D_-} \, .
\label{boundary to bulk propagator near ads2 B}
\eea
The solution at these times exhibits ``transient'' Aretakis behavior, since
\bea
\left. \pa^k_r G(v,r;v_0) \right|_{r=0} \propto (v-v_0)^{k+\D_-} \, ,
\label{transient aretakis behavior}
\eea
to leading order in $\k (v-v_0)$. Thus, this constitutes a symmetry argument for the ``transient'' Aretakis instability for near-extreme RN BHs. A translation of the statement (\ref{transient aretakis behavior}) to the full extreme RN geometry (eliminating $v_0$ by a time translation) would be that for advanced times $M^{-1} \hat{\k} \hat{v} \ll 1$ derivatives of the field behave as
\bea
\left. \pa^k_{\hat{r}}\phi \right|_{\hat{r}=0} \propto \hat{v}^{k+\D_-} \, ,
\label{transient aretakis behavior far coords}
\eea
and for times $M^{-1} \hat{\k} \hat{v} > 1$ the field exhibits exponential decay through QNMs as shown in equations (\ref{nads2 qnms l=0}),(\ref{nads2 qnms l>0}).

For more general cases (such as massive/charged fields, axisymmetric modes of rotating BHs, and probably also higher spin fields), it seems likely that a similar argument would go through.
For example, in extremal Kerr, for modes with complex $\Delta_- = 1/2 + i r$ with $r$ real, at late times, the solution may be written as \cite{Gralla:2017lto}
\bea
\phi \sim V^{\Delta_-} \sum_n V^{i \, \chi_n}  f_n(V R) \, ,
\label{late time behavior complex delta}
\eea
where $\chi_n$ are some real phases and $f_n$ are smooth. Thus if we transform to coordinates of the type (\ref{V and R}) (near-NHEK coordinates, in this example) these modes will exhibit a ``transient instability'' behavior for $\k v \ll 1$.

\section{Backreacted solution within Jackiw-Teitelboim theory}\label{backreacted solution}

What ``happens'' to a perturbed extremal BH? What is the endpoint of the Aretakis instability and how does the system evolve toward it? This was addressed in \cite{Murata:2013daa} by numerically solving the coupled Einstein-scalar equations for spherically symmetric massless, neutral scalar perturbations around extreme RN. The process was understood there as follows: the instability persists nonlinearly, as explained in section \ref{Introduction}; but for generic initial data, at some point backreaction kicks in, causing the BH to become slightly non-extremal, and thus small perturbations at late times eventually decay exponentially at a timescale corresponding to the final inverse temperature. This makes sense from an outside observer's point of view: perturbing a system at the absolute zero with some small but nonzero energy, it seems reasonable that its final state would be, at least in some effective sense, thermal. But how can this be connected to the near-horizon picture for the late-time behavior discussed in sections \ref{Instability from symmetry}, \ref{Transient behavior and quasinormal modes from symmetry}? As argued in \cite{Maldacena:1998uz}, pure \ads2 is incompatible with nonsingular finite energy excitations, when backreaction is taken into account. However, as was argued in \cite{Almheiri:2014cka} and refined in \cite{Maldacena:2016upp}, Jackiw-Teitelboim (JT) theory constitutes a universal model which allows the treatment of backreaction effects for \emph{nearly-\ads2} spacetimes. There has been much recent interest in the JT model, a 1+1 dimensional theory of dilaton gravity which we will discuss in this section. This was spurred most prominently by the realization \cite{Maldacena:2016hyu} that this theory captures the dynamics of the low-energy reparametrization mode of the Sachdev-Ye-Kitaev (SYK) model \cite{Kitaev talk},\cite{Sachdev:1992fk} and thus by the prospect of finding a new, solvable model of holographic duality between a (0+1) dimensional boundary and a (1+1) dimensional bulk. It also constitutes a controllable toy model to address questions of principle in quantum gravity, c.f. \cite{Maldacena:2017axo},\cite{Harlow:2018tqv}. In our context, JT theory will be useful since it allows a remarkably simple description of \emph{backreaction} effects in the near-horizon region of near-extremal BHs described by the nearly-\ads2 geometry; thereby we will be able to study the nonlinear evolution of near-extremal BHs at late times analytically.

At least to some degree of universality, it has recently been argued that the JT model constitutes a general description of the low energy perturbations of extremal BHs. This has been explicitly shown for spherically symmetric perturbations of large extremal RN-AdS BHs in \cite{Nayak:2018qej}\footnote{and non-spherically symmetric perturbations can presumably also be accounted for by a Kaluza-Klein reduction into massive fields in 2 dimensions, see recent discussion in \cite{Moitra:2018jqs}.} (see also \cite{Sarosi:2017ykf}). In this section we will be interested in the simplest case of spherically symmetric perturbations of a minimally coupled massless, neutral scalar field around extreme RN, and therefore we will study JT theory coupled to a massless neutral scalar in near-\ads2.

The JT theory is given by the action
\bea
S = \frac{1}{2} \int \sqrt{-g} \, d^2 x \, \, \Phi(R+2)+\Phi_\mathrm{bdy} \int_{\mathrm{bdy}} \sqrt{h} \, (K-1) \, du +S_{\mathrm{matter}} \, ,
\label{JT action}
\eea
where $K$ is the extrinsic curvature of the boundary, $\sqrt{h}$ is the induced metric on the boundary and $u$ is the proper time along it, $\Phi$ is a usually referred to as the dilaton, and $S_{\mathrm{matter}}$ is the bulk matter action. Importantly, we adopt the assumption of \cite{Almheiri:2014cka},\cite{Maldacena:2016upp} that the bulk matter Lagrangian is independent of $\p$, so there is no direct matter-dilaton coupling. $\p$ encodes the volume of the transverse $S^2$ of the original higher-dimensional geometry\footnote{In fact, a reduction from higher dimensions will result in another, topological, term $S_{\mathrm{Topo}} = \frac{1}{2} \Phi_0\left( \int R+2\int_{\mathrm{bdy}} K \right)$ (where $\Phi_0$ is a constant); but this term is just a constant in our case since the topology is fixed (and is actually equal to 0 since we consider the Lorentzian strip topology). $\Phi_0+\Phi$ is the total volume of the transverse $S^2$.}.
At large radii, the dilaton generically diverges and therefore the solution needs to be cut off at some radial distance $\sim \epsilon^{-1}$, where $\epsilon$ is a small parameter controlling this cutoff. The physical boundary is taken to be the (1-dimensional) timelike hypersurface where the value of $\p$ is fixed to some value $\Phi_{bdy}:=\Phi_r \epsilon^{-1}$, where $\Phi_r$ is a renormalized value for the dilaton, factoring out its scaling with $\epsilon$. 

Even though the theory (\ref{JT action}) is already quite simple, it turns out that it admits an alternative formulation which looks even simpler, and is particularly useful for the present application \cite{Maldacena:2016upp},\cite{Maldacena:2017axo}. $\p$ is an auxiliary (nondynamical) field, and it is simply integrated out exactly, setting the Ricci Scalar $R=-2$. This fixes the geometry, locally, to \ads2. The only remaining degree of freedom, then, is the boundary trajectory which is assumed to reside (radially) far from the horizon, so that the \ads2 throat is long. The gravitational part of the action then reduces to
\bea
S = \p_{\mathrm{bdy}} \int_{\mathrm{bdy}} \sqrt{h} \, (K-1) \, du  +S_{\mathrm{matter}}  \, .
\label{action reduction to bdy}
\eea
Parameterizing the boundary worldline using Poincar{\'e} coordinates $\{ T_b(u) , R_b(u) \}$ where $u$ is the proper time on the boundary, and using the relation $R^{-1}_b(u) \approx \e \dot{T}_b(u)$ for large $R_b$, the purely gravitational part of the action can be put in the form
\bea
S_{\mathrm{grav}} = -\p_r \int \{ T_b(u),u \} du   \, ,
\label{scwarzian action}
\eea
where 
\bea
\{ T_b(u),u \} = \frac{\dddot{T}_b}{\dot{T}_b} - \frac{3}{2} \left(\frac{\ddot{T}_b}{\dot{T}_b}\right)^{2} \, ,
\label{schwarzian def}
\eea
is the Schwarzian derivative, and $\, \dot{}=d/du \, $. (\ref{scwarzian action}) describes the purely gravitational (no matter fields of yet) dynamics of near-\ads2 spacetimes. It is \sl2r invariant; in particular in the vacuum case, a conserved energy may be defined:
\bea
- E = \p_r \{ T_b(u),u \} \, .
\label{energy pure gravity}
\eea

Now we would like to explicitly add bulk matter to this picture. As mentioned above, in the JT model matter fields propagate (locally) freely on an \ads2 background and are coupled to the gravitational degrees of freedom only through the boundary. 
We will assume for simplicity in this section Dirichlet boundary conditions which set the matter field to zero at the boundary, $\phi|_{\mathrm{bdy}}=0$ where $\phi$ is our bulk scalar field.\footnote{It is possible to modify the assumed boundary conditions. This may be desired since when reducing from a generic perturbed asymptotically flat BH we expect the field (and its derivative) to have some time dependent value on the boundary. The realistic, precise boundary conditions would be of Robin type, and time dependent; however Dirichlet, $\phi=0$ boundary conditions seem to be a good approximation for the low frequency modes which govern the late-time behavior. The solution found here, then, is expected to capture the late-time behavior even under more realistic boundary conditions.}
Following \cite{Maldacena:2016upp}, a particularly simple way to take into account the influence of matter fields is by keeping track of how the total energy E is distributed between the BH and the matter\footnote{This energy is conserved with the boundary conditions of choice here; more generally one would need to keep track of the energy ``leaking out'' to infinity.}. This may be done by reinstating the dilaton and writing down the ADM energy in its terms of its value, and the value of its normal derivative on the boundary, using a relation between total energy and variation of the action with respect to the boundary metric \cite{Balasubramanian:1999re}. Then one may use the EOM $\delta S/\delta g^{RR}=0$ in order to relate these quantities to the pressure component of the matter field's stress-energy $T_{RR}$ there. A detailed derivation (in Euclidean signature) is found in appendix A of \cite{Maldacena:2016upp}.
The result is a generalization of (\ref{energy pure gravity}) to non-vacuum solutions:
\bea
- E = \p_r \, \{ T_b(u),u \}  + \dot{T}_b \, R_b^3 \, T_{RR} \, ,
\label{energy balance equation}
\eea
which we find particularly useful for our application.

The details of $S_{\mathrm{matter}}$ are not universal---different matter content and other (symmetry) assumptions in the higher-dimentional parent theory, will generically yield different $S_{\mathrm{matter}}$. We will focus for simplicity in this paper on the case of spherically symmetric perturbations of a neutral, minimally coupled massless scalar field which will correspond to taking
\bea
S_{\mathrm{matter}} = - \frac{1}{2} \int \, \sqrt{-g} \, d^2 x \, (\nabla \phi)^2 \, .
\label{S matter}
\eea
Allowing, for example, non-spherically symmetric scalar perturbations will introduce a tower of massive fields. 
As noted above, a general feature of JT theory which crucially simplifies the analysis here is that $\p$ does not appear in $S_{\mathrm{matter}}$, and therefore does not couple directly to the matter fields at leading order in the strength of the perturbations; at leading order, gravity-matter coupling occurs only through the boundary degree of freedom.

We wish to solve equation (\ref{energy balance equation}) in the presence of a massless scalar field propagating in the bulk with Dirichlet boundary conditions. 
Assume that initial data is chosen at $V=0$ so that the outgoing component of the field is given, for $U>0$, by 
\bea
\phi_{\mathrm{outgoing}} = H_0 \, f(U) \, ,
\label{initial data def}
\eea
where $f$ is some arbitrary function determining the profile of the outgoing wave. We have factored out the amplitude of the perturbation $H_0$ (we will later assume it is small).
The boundary condition $\phi=0$ we impose at $R_b=\mathcal{O}(\e)$, then, determines the solution for the scalar field to be of the form
\bea
\phi = f(U)-f(V)+\D \phi(V) \, ,
\label{general scalar field solution}
\eea 
where $\D \phi = \mathcal{O}(\e)$. Specifically, near the boundary, we can expand
\bea
T_{RR} = \frac{2 H_0^2}{R^4} \left[(f'(T))^2 \, + \mathcal{O}(\e)\right] \, . 
\label{stress energy near boundary}
\eea
Equation (\ref{energy balance equation}), then, is given to leading order in $\e$ by
\bea
- E = \p_r \, \{ T_b(u),u \}  +   2 \e \, H_0^2 \, \dot{T}^2_b(u) \, (f'(T_b(u)))^2 \, .
\label{energy balance equation 2}
\eea
We wish to understand how the boundary trajectory is affected by radiation in the bulk, and how this affects the late-time behavior. 
To this end, we find it useful to revert at this stage to a Poincar{\'e} time parametrization $R_b(T)$, using $R_b^{-1} \approx \e \dot{T}_b$ 
to rewrite (\ref{energy balance equation 2}) as 
\bea
\left[\partial_{T} \left( \ln R_b \right)\right]^2 -2 \partial^2_{T} \left( \ln R_b \right) + \frac{(2 \pi)^2}{\b^2}  \e^2 \, R_b^2 =  -\frac{4 \e}{\p_r} H_0^2 (\partial_Tf)^2 \, ,
\label{energy balance equation 3}
\eea
where $\b=2\pi\sqrt{\p_r/(2E)}$. Equation (\ref{energy balance equation 3}) governs (together with the wave equation) the behavior of the BH+matter system for general near-horizon initial data. The choice of initial data enters through a specific choice of $f(T)$. In the vacuum case, the right hand side vanishes and (\ref{energy balance equation 3}) reduces to the statement of conservation of the Schwarzian $\{ T_b,u \}$.

For small perturbations, (\ref{energy balance equation 3}) may be solved perturbatively in $H_0$. Beginning by putting $H_0=0$, the solution has a particularly simple form:
\bea
R_{b,(0)}^{-1}(T) = \e \left[ \, 1- \, \left(\frac{\pi}{\b}\right)^2 \, T^2 \right] \, ,
\label{zeroth order solution R_b}
\eea
and we defined integration constants so as to get precisely $R^{min}_b = \e^{-1}$, where $R^{min}_b$ is the minimal value of the Poincar{\'e} radial coordinate along the boundary trajectory. There is also freedom to shift $T$ by a constant; for convenience we choose $R^{(0)}_b(T=0)=R^{min}_b$.
To perturb (\ref{zeroth order solution R_b}), it is convenient to use an ansatz of the form 
\bea
R_{b}^{-1}(T) = R_{b,(0)}^{-1}(T) + \frac{\e H_0^2}{\p_r} R_{b,(1)}^{-1}(T) \, .
\label{perturbation ansatz}
\eea
Plugging this into (\ref{energy balance equation 3}) and equating the $\mathcal{O}(\e H_0^2/\p_r)$ terms on both sides, we obtain an equation for the perturbation:
\bea
\partial_{T}^2 R_{b,(1)}^{-1} + 2\left[T^2- \left(\frac{\b}{\pi}\right)^2 \right]^{-1} \left( R_{b,(1)}^{-1} - T\partial_T R_{b,(1)}^{-1} \right) = \frac{2 \, \e \, \pi^2}{\b^2 } \left[T^2- \left(\frac{\b}{\pi}\right)^2 \right] \, (\partial_Tf)^2 \, .
\label{equation for delta}
\eea
This second order, non-homogeneous linear ordinary differential equation is integrable for arbitrary forcing term $f(T)$. The integration constants in the general solution are determined by the initial conditions, namely that for $V=0$ the solution and its derivative identify with the unperturbed solution (\ref{zeroth order solution R_b}).
Employing these conditions gives
\bea
R_{b,(1)}^{-1}(T) = 2 \e \left\{ \left(\frac{\pi}{\b} \, T - 1 \right)^2 \int_{\e}^{T} \, y \,(\partial_yf(y))^2 \, dy - T \int_{\e}^{T} \left(\frac{\pi}{\b} \, y-1\right)^2 (\partial_yf(y))^2 \, dy \right\} \, . \non \\ 
\label{general solution perturbed trajectory}
\eea
Notice that $R_{b,(1)}^{-1}$ is always negative in the relevant range: in our case, the boundary can only be pushed outwards, and the temperature increased.

To be more concrete, it will be interesting to consider the specific profile $f(T)$ which corresponds to the special---universal---near-horizon solution in the sense explained in preceding sections, namely 
\bea
\phi \simeq \frac{H_0}{V (2+V \, R)} = \frac{H_0}{2} \left( 1/V - 1/U \right) \, .
\label{universal late time solution near-horizon coordinates}
\eea
The $\propto 1/U$ piece in the above can be thought of as arising from near-horizon initial data corresponding to an outgoing wave and specified on, e.g., the surface $V=0$, and the $\propto 1/V$ piece can be thought of as its reflection off the boundary, given the boundary conditions we impose. It is interesting to note that the computation we perform is relevant also to the scenario where the Aretakis constants vanish, and the BH is perturbed from afar. As discussed in section \ref{Transient behavior and quasinormal modes from symmetry}, this amounts to considering a solution for $\phi$ which is proportional to the boundary-to-bulk propagator, just with an amplitude suppressed by a factor $\sim \e^{-\D_-}=\e$ since the point of emission/perturbation resides at large $R$,
and this propagator is of the exact same form (\ref{universal late time solution near-horizon coordinates}). Interestingly, this particular suppression of the amplitude with $\e$ will in fact pop up in the below analysis as a consistency condition for our perturbative analysis. Another important point to note is that, as in (\ref{general scalar field solution}), there will be corrections to (\ref{universal late time solution near-horizon coordinates}) due to the modification of the boundary trajectory---namely, the reflected part will be $\propto 1/V + \mathcal{O}(\e)$. For the trajectory computation, at leading order we can neglect it, but for example for the field's late time behavior this will be important to keep in mind. The above discussion amounts to simply plugging $f(T) = -(2 T)^{-1}$ in (\ref{general solution perturbed trajectory}). The solution then is given, to leading order in $\e$, by
\bea
R_{b,(1)}^{-1}(T) =  - \frac{ \Theta (T-\e)}{6  \e^2}  \left( T -\frac{3\e}{2} + \frac{\e^3}{2T^2} \right)  \, ,
\label{universal solution delta}
\eea
and note that the two rightmost terms in the brackets above are important only at early times $T \sim \e$.
This result, combined with (\ref{perturbation ansatz}) is telling us that in order for the perturbative computation to be consistent in the case (\ref{universal late time solution near-horizon coordinates}), it is mandatory to take the amplitude of the scalar field $\phi$ to scale no stronger than $\e$. Physically we can interpret this as the incompatibility of the near-horizon picture with too strong an amplitude, realized in \cite{Maldacena:1998uz}: it is impossible to squeeze perturbations of higher amplitude in the near-horizon region and still keep them outside the event horizon.\footnote{This criterion on the amplitude may be related to the distinction in \cite{Murata:2013daa} between ``degenerate apparent horizon'' and ``first order mass perturbation'' types of initial data considered there.}
Finally, it is also interesting to note that the extra factor of $\e$ we are forced to suppress the amplitude with arises naturally when considering the case of vanishing Aretakis constant---indeed, the far-near Green's function (\ref{boundary to bulk propagator}) (in other words, the solution in the near-horizon region arising from an instantaneous, localized perturbation in the far region) is precisely of the form (\ref{universal late time solution}), only suppressed with a factor of $\e$ as discussed above and in section \ref{Transient behavior and quasinormal modes from symmetry}. 

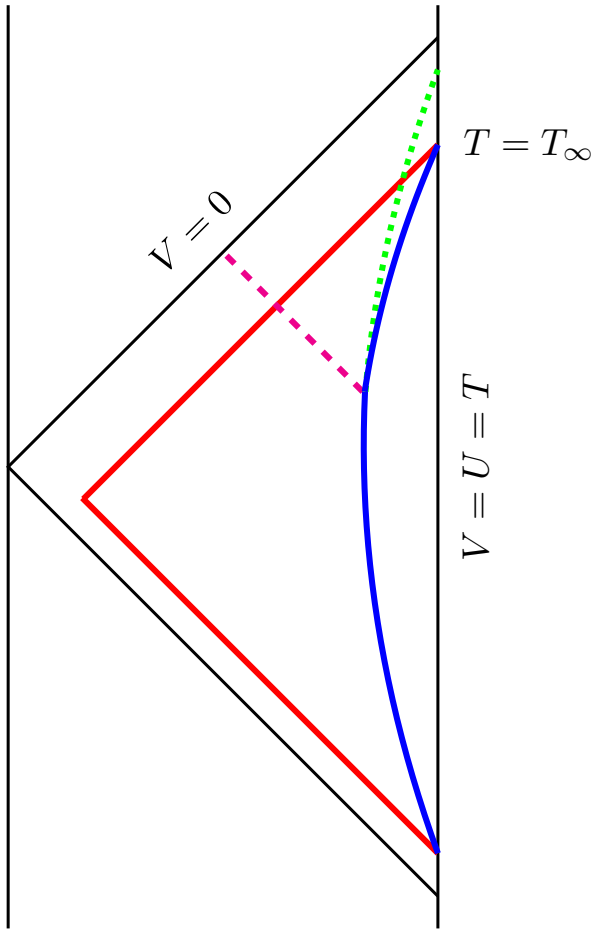
\begin{figure}[h!]
	\centering
	
	\resizebox{0.65\textwidth}{!}
	{
	
\begin{tikzpicture}
\draw[thick] (0,-0.3) -- (0, 8.3);
\draw[thick] (-4,-0.3) -- (-4, 8.3);

\draw[red, ultra thick] (-3.3, 3.7) to (0,7);
\draw[red, ultra thick] (-3.31, 3.71) to (0,0.4);

\draw[ultra thick, magenta,dashed] (-0.7,4.7) to (-2, 6);

\draw[thick] (0,0) to (-4, 4) to (0,8);

\draw (0.85,7.0) node[rotate=0]{\footnotesize{$T=T_\infty$}};
\draw (-5.5,4) node[rotate=0]{\footnotesize{$$}};

\draw (-2.3, 6.2) node[rotate=45]{\footnotesize{$V=0$}};
\draw (+0.35, 4) node[rotate=90]{\footnotesize{$V=U=T$}};
\draw[blue, ultra thick, domain=0.4:4.7][rotate=90] plot (\x, {-(0.0478)*(\x-0.4)*(\x-8)});
\draw[green, ultra thick, dotted, domain=4.7:7.7][rotate=90] plot (\x, {-(0.0478)*(\x)*(\x-7.7)});
\draw[blue, ultra thick, domain=4.7:7][rotate=90] plot (\x, {-(0.063)*(\x)*(\x-7)});

\end{tikzpicture}

	}
	
	\caption{Penrose diagram illustrating the near-horizon computation of late-time behavior including the effects of backreaction. The physical boundary, in solid blue, is promoted to a degree of freedom encoding the black hole's (small) deviation from extremality. The red solid lines are the future and past horizons. The magenta dashed line depicts a possible initial data surface, or an advanced time in which the BH is perturbed from afar. 
	The dotted green line shows the later part of the boundary trajectory had the scalar field not been excited. The late-time behavior of fields, as in fig. \ref{figure2}, is determined by analyzing the neighborhood of the point $T=T_\infty$.}
	\label{figure3}
\end{figure}

We will explicitly scale the amplitude, then, as $H_0:=\e h_0$ where $h_0=\mathcal{O}(\e^0)$. The full solution for the trajectory is now given by 
\bea
R_{b}^{-1}(T) = \e \left[1- \left( \frac{\pi}{\b}\right)^2 \, T^2  - \Theta (T-\e) \, \frac{ h_0^2}{6  \, \p_r} \left( T -\frac{3\e}{2} + \frac{\e^3}{2T^2} \right)  \right] \, .
\label{full solution trajectory}
\eea
In particular, as a consequence of the backreaction there will be a shift in the position of future timelike infinity (located at the intersection of the physical boundary and the boundary of global \ads2---see figure \ref{figure3}). It will now be located at 
\bea
T_{\infty} = \frac{\b}{\pi}\left( \sqrt{1+\xi^2} - \xi \right) \approx \frac{\b}{\pi} (1-\xi) \, ,
\label{T infinity}
\eea
where 
\bea
\xi = \frac{\b h_0^2}{12 \pi \p_r} \, .
\label{xi def}
\eea
In particular, $T_{\infty}$ decreases as the perturbation's amplitude $h_0$ increases. Importantly, the value of $T_\infty$ determines the position of the event horizon as the latter is determined simply by the latest right-moving null ray reaching $T_\infty$. Figure (\ref{figure3}) depicts schematically the backreacted near-horizon computation, and in particular the boundary's motion and the event horizons. 


Now that we have computed the boundary trajectory and the location of the backreacted future event horizon, we can proceed to determine the late-time behavior of the scalar perturbations. It is worth stressing again that in the bulk we are simply solving the wave equation on (a patch of) a fixed \ads2 geometry. We then only need to \emph{evaluate} the field along the correct curves, which correspond to the backreacted event horizon and boundary. The problem of determining the late-time behavior on the horizon now reduces, essentially, to that already solved in section \ref{Transient behavior and quasinormal modes from symmetry}. 
In particular, the solution will be given by equation (\ref{general scalar field solution}), where here we will include the correction $\Delta \phi (V)$ of equation (\ref{general scalar field solution}) to get the correct value of the field near the boundary.\footnote{This correction is negligible when computing the boundary trajectory, equations (\ref{energy balance equation}),(\ref{stress energy near boundary}).} 
In fact, with our choice of boundary conditions, the general solution (\ref{general scalar field solution}) is simply corrected to 
\bea
\phi = f(U)-f(V+2 R_{b}^{-1}(V)) \, ,
\label{scalar field corrected solution - general}
\eea
where $R_{b}^{-1}$ is given by (\ref{general solution perturbed trajectory}).
For concreteness and since it is especially important (due to its universality) we will restrict to the case (\ref{universal late time solution near-horizon coordinates}) (the same method in principle will apply for the general solution, (\ref{general solution perturbed trajectory})). The corrected solution then is
\bea
\phi = \frac{\e h_0}{2} \left[ \frac{1}{V+ 2 R_{b}^{-1}(V)} - \frac{1}{U}  \right] \, ,
\label{scalar field corrected solution}
\eea
where $R_{b}^{-1}$ is given by (\ref{full solution trajectory}) to leading order in $\e$.
The exterior of the BH---the part which needs to be matched, along the boundary, to a near-extreme RN geometry---is determined by the boundary trajectory. In our case it is a near-\ads2 patch which begins at the Poincar{\'e} time $T_{-\infty} = -\beta/\pi$ and ends at $T_{\infty} = \beta (1-\xi)/\pi$. In a fashion very similar to the analysis of section \ref{Transient behavior and quasinormal modes from symmetry}, we will use the mapping
\bea
U &=& -\frac{\xi \b}{2 \pi}+\frac{\b}{\pi}(1-\xi/2)\tanh\left[ \frac{\k(v-v_0)}{2} - \frac{1}{2} \ln\frac{r}{r+2\k} \right] \, , \non \\
V &=& -\frac{\xi \b}{2 \pi}+\frac{\b}{\pi}(1-\xi/2)\tanh\left[ \frac{\k(v-v_0)}{2} \right] \, ,
\label{generalized near-ads2 to ads mapping}
\eea
where $v_0$ is a free parameter, which takes part of \ads2 in double-null coordinates with metric (\ref{near-horizon metric in double-null coordinates}) to a near-\ads2 patch in ingoing Eddington-Finkelstein coordinates,
\bea
\frac{ds^2}{M^2} = -r(r+2\kappa) dv^2 +2dr dv + d\Omega^2 \, ,
\label{near ads2 in ingoin EF coords}
\eea
to study the late-time behavior, as depicted in figure \ref{figure3}. Expanding (\ref{generalized near-ads2 to ads mapping}) near $T_{\infty}$ gives
\bea
U-T_{\infty} &=& -\frac{\b}{\pi}(2-\xi) \frac{r}{r+2\k} e^{-\k(v-v_0)} \, , \non \\
V-T_{\infty} &=& -\frac{\b}{\pi}(2-\xi) e^{-\k(v-v_0)} \, ,
\label{generalized near-ads2 to ads mapping, expanded}
\eea
which reproduces (\ref{approx near future point}) up to an overall constant which will be insignificant for the analysis here, and in particular may be absorbed into the factor $e^{-\k v_0}$ by a redefinition of $v_0$. We may then proceed to expand the solution as in equation (\ref{nads2 qnms l=0}) of section \ref{Transient behavior and quasinormal modes from symmetry}, obtaining a sum over quasinormal modes which decay exponentially, as $e^{-\k v}$, at late times. Note that the QNM amplitudes are given by derivatives of the functions $f$, $g$ which appear in equation (\ref{scalar field corrected solution}).

It will be interesting to use the analysis of this section to study in detail the field's behavior along the boundary at all times (including early times), and compare with expectations from holography. In principle this may be done by writing (\ref{scalar field corrected solution}) near the boundary as
\bea
\phi \simeq  \frac{\e h_0 }{(V+ 2 R_{b}^{-1}(V))^2} \, (R^{-1}-R_b^{-1}(V)) + \mathcal{O}(R^{-1}-R_b^{-1}(V))^2   \, .
\label{corrected solution near the boundary}
\eea
The first term on the right hand side of (\ref{corrected solution near the boundary}) determines the time dependence (upon translation to boundary time) of the field at the boundary.\footnote{The solution vanishes on the boundary by construction, as required by the boundary conditions---but its normal derivative, for example, does not.} Note that while equation (\ref{corrected solution near the boundary}) is exact in $\e$, when we plug in the solution (\ref{full solution trajectory}), which is correct only to leading order, we get the late-time behavior only up to next-to-leading order in $\e$. The late-time decay will be exponential, as in the bulk, but there may be interesting time dependence at earlier times and at subleading orders in $\e$. The (transient instability) behavior on the horizon at intermediate times is given by the corresponding expression from section \ref{Transient behavior and quasinormal modes from symmetry}.

But the above analysis gives the late-time behavior only in terms of near-horizon quantities, and in particular $\k$ was introduced in (\ref{generalized near-ads2 to ads mapping}) as an arbitrary parameter. In the higher-dimensional setting, we are considering a situation where a near-extremal BH is perturbed with energy comparable to its own energy above extremality. Since the perturbation is spherically symmetric, the far-region geometry will settle down to near-extreme RN at late times, so in order to translate from near-region to far-region language we only need to determine what is the temperature/surface gravity of the new, resulting BH (we assume that the temperature of the old one is $\hat{\k}$ as in section \ref{Transient behavior and quasinormal modes from symmetry}). The change in temperature of the BH can be inferred from the shift in $T_\infty$, equation (\ref{T infinity}). The perturbed trajectory traverses a shorter Poincar{\'e} time than the original one, by a factor of $(1-\frac{\b h_0^2}{12 \pi \p_r})$, but must traverse the same (very large) amount of boundary/proper time as the unperturbed trajectory. This means that $\frac{d}{du}$ must pick up an inverse factor to the above, and therefore the boundary energy, proportional to the Schwarzian, will pick up a factor of $\sim (1+\frac{ \b h_0^2}{6 \pi \p_r})$. But given that the surface gravity is $\sqrt{1-Q^2/M^2}$ for $M\simeq Q$, the perturbed temperature/surface gravity will be enhanced by a factor of $(1+\frac{\b h_0^2}{12 \pi \p_r})$.

It is tempting to go further and identify $(2\pi \e)/\b$ as the initial surface gravity $\hat{\k}$ and $h_0^2/\p_r$ as the amplitude squared $A^2$ of the spherically symmetric perturbation in appropriate units. This yields, for the surface gravity of the new BH,
\bea
\hat{\k}' \simeq \hat{\k}+\frac{A^2}{12 M},
\label{new surface gravity}
\eea
In particular, the late-time behavior will be given by equation (\ref{nads2 qnms l=0}), but now translating to the far-region quantities one would need to replace $\hat{\k}$ with $\hat{\k}'$ of equation (\ref{new surface gravity}), yielding for the QNM frequencies governing the behavior of the new ringing-down BH,
\bea
\hat{\omega}_n = -i\hat{\k}'n = -i  (\hat{\k}+\frac{A^2}{12 M}) n  \, .
\label{new QNM frequencies}
\eea
In particular, the field and its derivatives decay on the horizon at late times as
\bea
\phi \propto e^{-\hat{\k}' \hat{v}} \, ,
\label{new late-time behavior}
\eea
with $\hat{\k}'$ given by (\ref{new surface gravity}).
It would be interesting to verify in detail this matching to far-region quantities.

Note here that sometimes it makes more practical sense to leave results in the resummed form (\ref{l=0 solution nads2}) than expand to the QNM-like form (\ref{nads2 qnms l=0}). Indeed, in some situations one finds that near-extremal QNMs are explicitly summable analytically, e.g. \cite{Hadar:2014dpa},\cite{Gralla:2016sxp}, but this should be of no surprise since the QNM sum ``originates'' from a resummed expression as in (\ref{l=0 solution nads2}) to begin with.

\section*{Acknowledgements}

I am grateful to Joan Camps, Hadi Godazgar, Michal Heller, Achilleas Porfyriadis, and Andy Strominger for useful conversations and to Sam Gralla, Alex Lupsasca, and Harvey Reall for helpful comments on the manuscript. I gratefully acknowledge support by the Jacob Goldfield Postdoctoral Support Fund and by the Max Planck Gesellschaft through the Gravitation and Black Hole Theory Independent Research Group.

\end{document}